# Evaluating experiences in a digital nutrition education program for people with multiple sclerosis: a qualitative study

Russell RD[a], He J[a], Black LJ[a,b], Begley A[a]
[a]Curtin School of Population Health, Curtin University, Perth, Australia
[b]Institute for Physical Activity and Nutrition (IPAN), School of Exercise and Nutrition Sciences, Deakin University, Melbourne, Australia

**Corresponding author:**
Rebecca Russell
Rebecca.Russell@curtin.edu.au

**Data availability statement**

The data that support the findings of this study are available from the corresponding author upon reasonable request.

**Conflicts of interest**

No authors declare no conflicts of interest

**Funding**

This research was supported by MSWA and an MS Australia Incubator Grant (#21-1-072). LJB is supported by MSWA and an MS Australia Postdoctoral Fellowship. RDR is supported by MSWA. Funders had no role in the design, analysis, or writing of this article.

**Authorship**
RDR: Conceptualisation, Investigation, Writing – Original Draft, Writing – Review & Editing, Project administration, Funding Acquisition, Formal Analysis, Methodology, Project Administration
JE: Formal Analysis, Data Curation, Project Administration, Validation, Writing - Original Draft
LJB: Conceptualisation, Methodology, Writing – Review & Editing, Supervision, Funding Acquisition, Validation
AB: Conceptualisation, Methodology, Writing – Review & Editing, Supervision, Funding Acquisition, Formal Analysis, Validation

**Acknowledgements**

We thank the MS community for their involvement in co-designing "Eating Well with MS", and the participants who took part in the interviews.

## Abstract

**Background**

Multiple sclerosis (MS) is a complex immune-mediated disease with no currently known cure. There is growing evidence to support the role of diet in reducing some of the symptoms and disease progression in MS, and we previously developed and tested the feasibility of a digital nutrition education program for people with MS.

**Objective**

The aim of this study was to explore factors that influenced engagement in the digital nutrition education program, including features influencing capability, opportunity, and motivation to change their dietary behaviours.

**Methods**

Semi-structured interviews were conducted with people who MS who completed some or all of the program, until data saturation was reached. Interviews were analysed inductively using thematic analysis. Themes were deductively mapped against the COM-B behaviour change model.

**Results**

16 interviews were conducted with participants who completed all (n=10) or some of the program (n=6). Four themes emerged: 1) Acquiring and validating nutrition knowledge; 2) Influence of time and social support; 3) Getting in early to improve health; and 4) Accounting for food literacy experiences.

**Discussion**

This is the first online nutrition program with suitable behavioural supports for people with MS. It highlights the importance of disease-specific and evidence-based nutrition education to support people with MS to make dietary changes. Acquiring nutrition knowledge, coupled with practical support mechanisms such as recipe booklets and goal-setting, emerged as crucial for facilitating engagement with the program.

**Conclusions**

When designing education programs for people with MS and other neurological conditions, healthcare professionals and program designers should consider flexible delivery and building peer support to address the needs and challenges faced by participants.

**Patient or Public Contribution**

Members of the MS Nutrition Research Program Stakeholder Reference Group, which comprises of people with MS and MS health professionals, provided input during the development of the nutrition education program and study design stages.

**Practitioner Points**

When designing and developing nutrition education for participants with neurological conditions, practitioners should:
- Ensure the education content is specifically tailored to suit the population
- Enlist suitable behavioural supports to facilitate dietary changes
- Encourage family or peer support to facilitate participation and engagement

## 1. Introduction

Multiple sclerosis (MS) is an immune-mediated disease characterised by inflammation and scarring of myelin sheaths in the central nervous system [1]. The pathological process gives rise to a wide range of symptoms, including sensory disturbances, visual impairments, motor deficits, vertigo, speech difficulties, and unpredictable prognosis [2]. MS is most commonly diagnosed in females 30-40 years of age and currently, there is no cure [2].

Approximately 40% of people newly diagnosed with MS make dietary changes [3]. Although there is growing evidence to support the role of diet in reducing some symptoms, disease progression, and improving quality of life among people with MS [4-7], there is not yet a specific therapeutic diet recommended for people with MS. The Internet is a primary source of dietary information for people with MS [8]; however, information retrieved is conflicting [9]. Diets marketed to people with MS vary greatly in composition, including restrictive diets that remove entire food groups [9], potentially risking nutrient deficiencies - some of which can exacerbate MS symptoms [10]. Conflicting information results in confusion about what dietary changes to make and where to seek dietary advice [11]. Given the complexity and variability of symptoms, there is a need for targeted nutrition education to empower people with MS to make informed dietary choices to improve their health and well-being. In our previous research, people with MS have told us that they want a nutrition education program to help alleviate the confusion surrounding diet [11].

To meet the needs of the MS community, we co-designed a nutrition education program with theoretical frameworks and behaviour change techniques (BCTs, [12]) to support dietary behaviour change in this population. Details of the program development have been published [13]. Briefly, the asynchronous digital program, titled "Eating Well with MS", adopted a multimodal delivery that incorporated text, interactive graphics, videos, and printed resources such as an activity book, recipe booklets, and information brochures. The education program included seven modules (Table 1). We tested the feasibility using a single-arm pre-post design (n=67 people with MS). All feasibility metrics indicated success (recruitment targets, participant completion, satisfaction, and limited efficacy testing) (manuscript under review). To date, we have assessed the program using quantitative methods; however, qualitative research can provide deep insights into participants' experiences to support further improvements of the program [14].

Table 1. "Eating Well with MS" program objectives and weekly modules

| Program objectives |
|---|

| To provide participants with the knowledge and skills to: | 1) manage their symptoms through healthy eating |
| --- | --- |
| | 2) assess the quality of their eating habits |
| | 3) select, prepare, and cook healthy meals |
| | 4) judge the credibility of special diets that are marketed to people with MS |
| | 5) explain how researchers develop evidence in the field of nutrition and MS |
| | Module title |
| Module 1 | Program introduction |
| Module 2 | Healthy eating is important for people with MS |
| Module 3 | Personalising your eating habits |
| Module 4 | Making changes to your eating habits |
| Module 5 | Understanding the diets that are marketed to people with MS |
| Module 6 | Understanding the research on diet and MS |
| Module 7 | Summary |

MS, multiple sclerosis.

Qualitative methodologies are well-suited to exploring the complexities of food-related behaviour and can facilitate the collection of authentic experiences to interpret the quantitative data from our feasibility study [14]. In conjunction with a qualitative approach, the use of a theoretical framework provides a systematic lens to analyse and understand individual experiences. The COM-B behaviour change model is well-established in the behavioural sciences and recognises that for behaviour change to occur, an individual needs to have the necessary capabilities (mental and physical), the right opportunities (external and environmental factors), and sufficient motivation (reflexive and automatic) to engage in the desired behaviour [15]. Given the multifaceted nature of MS and the challenges faced by people with MS, the COM-B model provides a suitable perspective to explore factors that may influence their ability to engage in a digital nutrition education program and to make dietary behaviour changes. Therefore, the aim of this study was to explore factors that influenced participant engagement in a digital nutrition education program for people with MS. The objectives were to: 1) identify program features influencing capability, opportunity, and motivation to change dietary behaviours; 2) explore participants' preferences for program features; and 3) identify how the program might be improved.

## 2. Methods

### 2.1 Study design

Using a qualitative descriptive approach, we explored the experiences of people with MS who participated in "Eating Well with MS" [16]. All participants provided online informed consent via Qualtrics (Version 2020, Provo, UT). The study was approved by the [Blinded for peer-review] University Human Research Ethics Committee (HRE2022-0020) and adhered to the Consolidated Criteria for Reporting Qualitative Research (Appendix A [17]).

**2.2 Participant recruitment**

Participants were recruited from the "Eating Well with MS" feasibility study (September-November 2022). All 67 participants were emailed information about the study and invited to participate. One potential participant withdrew before interview due to a family bereavement, and another expressed interest via email but did not complete the booking form. Purposeful sampling was used to identify and send a follow-up invite to people who had not completed the entire program to mitigate completion bias.

**2.3 Data collection**

Semi-structured, one-on-one interviews (audio-only) were conducted in December 2022. The interview guide contained 19-24 questions (depending on how much of "Eating Well with MS" the participant completed) (Appendix B). The questions were based on the COM-B model and preferred program features, and were refined through discussion by the research team. The interviews were conducted by Author1 via telephone or Microsoft Teams (version 1.6.00.19353, Microsoft Corporation; audio-only). Before commencing the interview, participants were reminded of the purpose and anticipated duration, and confirmed their consent. Rapport was established before beginning the interviews through discussions about the weather, holiday plans, and family. Interviews were recorded and transcribed verbatim and transcripts were emailed to participants to edit and/or add further insights, to ensure that the transcribed data accurately represented their perspectives (member checking) [18]. No transcripts were returned. Once thematic saturation was reached (no new codes emerging) [19], two additional interviews were conducted to confirm saturation.

**2.4 Data analysis**

Transcript data were uploaded to NVivo (version 12.6.0, QSR International Pty Ltd, Australia) for data management. The transcripts were analysed by Author2 following Braun and Clark's reflective thematic analysis [20]. Inductive analysis began with Author2 becoming immersed in the data (listening to each audio file multiple times). Transcripts were read line-by-line to identify significant sections of text, which were organised into groups and represented by codes. Iterative reviews of the codes revealed potential subthemes that emerged from the participants' responses, and a thematic map was created to visualise the relationships between the codes. The emerging themes were discussed with the research team to ensure methodological rigour and the trustworthiness of the findings [21]. As near-final themes emerged, deductive analysis was conducted to map the themes against the COM-B model (psychological and physical capabilities, environmental and social opportunities, and reflective and automatic motivations) [15]. Themes were reviewed and finalised after an in-depth discussion between Author2 and Author1 and confirmed by Author4 (an experienced qualitative researcher).

## 3. Results

### 3.1 Participant characteristics

We interviewed sixteen participants (Table 2). The age range was 34-71 years, nearly all were female (n=15). The majority had relapsing-remitting MS and the median time since diagnosis was nine years (range 10 months to 38 years). Less than half were currently employed (44%) and had attained a Bachelor's degree or higher (44%). Ten participants had completed all the modules of "Eating Well with MS" and six had completed half or less. The interview duration ranged from 22 to 71 min (median 34 min).

Table 2. Participant characteristics (n=16)

| | |
|---|---|
| **Sex, n (%)** | |
|     Female | 15 (94) |
|     Male | 1 (6) |
| Age (years), mean (SD) | 50 (12) |
| Time since diagnosis (years), median (IQR) | 9 (13) |
| **Type of MS, n (%)** | |
|     Relapsing-remitting | 11 (69) |
|     Primary-progressive or secondary-progressive | 4 (25) |
|     Unsure | 1 (6) |
| **Employment status, n (%)** | |
|     Employed | 7 (44) |
|     Retired or disability pension | 5 (31) |
|     Unemployed | 3 (19) |
|     Volunteering | 1 (6) |
| **Highest education attained, n (%)** | |
|     Bachelor's degree or postgraduate degree | 7 (44) |
|     Year 12 or equivalent | 5 (31) |
|     TAFE technical certificate or Diploma | 3 (19) |
|     Trade/apprenticeship | 1 (6) |

IQR, interquartile range; MS, multiple sclerosis; SD, standard deviation; TAFE, Technical and Further Education

### 3.2 Themes

Four themes emerged: 1) Acquiring and validating nutrition knowledge; 2) Influence of time and social support; 3) Getting in early to improve health; and 4) Accounting for food literacy experiences. Pseudonym, sex, time since diagnosis, and stage of program completion are detailed after each quote.

**3.2.1 Theme 1: Acquiring and validating nutrition knowledge**

For participants who had given some time and thought about their dietary habits, the program served as confirmation that they were eating well.

> "The main reason for signing up would be just validation of what conclusions I had come to on my own…for me the most valuable thing was just you're on track." Joy, female, 5 years since diagnosis, completed entire program.

Others shared that the program was a reminder or refresher to focus on eating well and prompted them to make dietary changes for themselves and those around them (e.g., their children and partners).

> "This was quite good because it did confirm that I should just go back to eating vegetables, eating fruit, eating fibre, and drinking water." Trudy, female, 6 years since diagnosis, completed entire program.

They described their appreciation of how the content was presented; in particular, they noted that information about specific diets was delivered in an unbiased way. For those who claimed to be following specific diets, such as the Overcoming MS diet or the Wahls Protocol, they described feeling that they could now make informed decisions about their dietary choices – such as increasing fish, dairy, and/or red meat intake (if they had previously excluded those foods).

> "It was good to see the program and go 'oh yeah, how about we look at some science, let's go with that; it's evidence-based'. I have been having problems with my iron and there was a piece that was talking about the proteins and the balances and things like that. I now have a little bit of turkey and a little bit of the low-fat red meat." Vicky, female, 4 since diagnosis, completed most of the program.

One participant who did not complete the program expressed that the content was not specific enough for them. While the information aligned with what their neurologist had told them, they said they were expecting more in-depth information about specific foods or ingredients (such as turmeric).

> "I found it all a bit too basic and not MS-centric enough...it's not really MS-based. It didn't really have that link for specific information for people with MS." Robin, male, 6 years since diagnosis, completed Modules 1-2.

The discussions also highlighted how to improve the program by personalising the content, by including optional information for people with higher baseline nutritional knowledge.

"Maybe if there was an advanced stream as well, for people who have a bit more knowledge with diet lifestyle." Stella, female, 1 year since diagnosis, completed half the program.

**3.2.2 Theme 2: Influence of time and social support**

The program was described as containing a comfortable amount of content to work through. Even participants who did not finish the whole program agreed that the amount of content and delivery pace were good, but unexpected circumstances prevented them from finishing. These circumstances included family illness, bereavement, work demands, and juggling family life.

"We had COVID, I work full time, I've got two little kids. I just didn't expect that I wouldn't be able to do it, but other things came up." Clarice, female, 1 year since diagnosis, completed Modules 1-3.

All participants except one described their intention to complete the full program, "I thought I would be able to do the whole thing really easily but...I went away a few times [and] work was a little bit busy." (Vicky, female, 4 years since diagnosis, completed most of the program). The participants who completed all the program generally described themselves as having more free time, for example, those who were retired.

"I'm retired, so it was very easy for me to get through it." Florence, female, 38 years since diagnosis, completed entire program.

The participants' social support circumstances also appeared to influence education program engagement. Some participants described family support from as being helpful to keep them on track. They discussed the program with family or friends and acknowledged that the program benefitted the eating habits of those around them, and recommended that a future program should involve family members of people with MS.

"My daughter actually asked me, 'Can we try something? You know, instead of this [meal], can we do this [recipe from Eating Well with MS] instead?' So, she's influenced me. She has a meal with me three times a week." Hilda, female, 7 years since diagnosis, completed entire program.

There was also value in sharing experiences with peers through the program discussion boards. The online social interactions helped participants feel less isolated and provided a space to swap ideas.

"I know that there's other people out there doing the same thing and I can go into the [discussion] forum and see what everybody else is doing…Knowing they're in the same sort of spot that I am really helped me." Florence, female, 38 years since diagnosis, completed entire program.

However, they shared their disappointment with the lack of conversation on the discussion boards, particularly as the weeks progressed. They were not sure how to encourage others to post, but that it would be valuable if the online conversations continued throughout the program and beyond in a private social media group.

"I don't know how you can encourage people to do it [post on discussion boards]…I was thinking along the lines of the Facebook-type thing; if it was available all the time you could go on there." Florence, female, 38 years since diagnosis, completed entire program.

### 3.2.3 Theme 3: Getting in early to improve health

The participants' motivation to engage in the program appeared to be shaped by their health concerns, such as fatigue, weight, and other health conditions, and prioritisation of their health. The program was described as a catalyst to improve their food choices, "[It] jolted me more into thinking I really should get back to eating better" (Stacy, female, 29 years since diagnosis, completed entire program) or to improve their weight.

"I've been overweight for a while and my husband is overweight…it's probably time we start changing up some things a bit and improving what we eat, so this [program] was an opportunity to start at the beginning and learn something." Dianne, a female, 11 years since diagnosis, completed the entire program.

Regardless of how long ago they were diagnosed, the program was described as helpful, and would have been particularly beneficial when they were newly diagnosed. In reflecting on their own experiences, they described how this would have helped them avoid searching through "lots of different, conflicting information" (Clarice, female, 1 year since diagnosis, completed Modules 1-3), which was described as overwhelming.

"It would have been amazing to have something like this when I was diagnosed. I could imagine how doing something like this for a newly diagnosed person would be helpful." Renae, female, 34 years since diagnosis, completed Modules 1-2.

"I really thought that program was great. Say my neurologist had said, 'Hey, you've been recently diagnosed; there's a program that's being run every few months with this University'…I would have signed up for it in a heartbeat…I just wanted someone who I felt like I could trust and who was a credible source to tell me [what to eat]" Joy, female, 5 years since diagnosis, completed entire program.

### 3.3.4 Theme 4: Accounting for food literacy experiences

Participants described their interactions and engagement with the program resources (e.g., activity book and recipe booklets), and some activities, such as meal planning, goal setting, modifying recipes, and reading food labels to make heathier choices. Overall, they found the resources and activities to be useful.

"The books were very, very helpful. The [label reading] card I use all the time." Florence, female, 38 years since diagnosis, completed entire program.

"I liked one of the recipes - that's gone into my diet [now]." Ester, female, 13 years since diagnosis, completed entire program.

There was variability in the use of food literacy skills, such as planning weekly meals and writing shopping lists, if they had not done much before. For others, the program prompted them to get back into planning meals to improve their diets. Participants described how the resources supported them in making changes to their dietary habits.

"I do think that [meal planning activity] is really useful for people that may not do it, for them to get ideas of how to get in your good food at the right time and prepare it prior." Vicky, female, 4 years since diagnosis, completed most of the program.

Occasionally, participants expressed that meal planning did not fit in with how they decided what to cook due to financial constraints.

"We're low-income, so quite often I'm looking for the markdowns and then just winging it [the meals] based on what could be available [in the shops]. Having some planning of my meals and then that rolling onto the shopping trolley - that was a big change; I'm not used to doing it that way." Dianne, female, 11 years since diagnosis, completed entire program.

## 4. Discussion and Conclusion
## 4.1 Discussion

This qualitative study describes the experiences of participants who completed some or all of a digital nutrition education program designed for people with MS in Australia. We aimed to explore the factors that influenced engagement in the program, including features that influenced capability, opportunity, and motivation to change dietary behaviours; preferences for program features; and program improvements. The themes identified factors that may contribute to dietary behaviour change in people with MS and engagement with a digital nutrition education program, such as acquiring or validating nutrition knowledge and skills (psychological and physical capability), the influence of life circumstances (e.g., retirement had a positive influence, whereas juggling multiple commitments had a negative influence) (physical and social opportunity), prioritising health concerns (reflective motivation), and varying food literacy skills (psychological and physical capability). This study provides valuable insights for the design and implementation of digital programs to promote and support dietary behaviour change in people with MS.

The nutrition education program was used by participants to acquire new nutrition knowledge or to validate their current understanding of nutrition for MS. Obtaining new knowledge can be a stimulus for adopting new lifestyle habits among people with MS [22], and an online health program for people with MS was reported to improve knowledge and self-reported dietary behaviour change [23]. Our finding that participants enrolled in our program to obtain new knowledge or validate their current knowledge is supported by other qualitative research involving people with MS who had completed an online MS course [24]. The study explored the motivations, behaviours, and expectations of web-based health information seeking and found that people with MS enrolled in the online course to either learn new information or consolidate their existing knowledge [24]. Discovering new information was a motivator for engaging with online health information [24]. In agreement with our findings, interviews with people with MS after they completed another online MS course revealed that some participants wanted more in-depth information and that some of the content was not specific enough to their needs [25]. It was evident that while most participants valued evidence-based information; however, given the expanse of non-evidence-based information available online for people with MS [9], some participants wanted more detail in less evidence-based areas they had read about for MS, e.g., turmeric. Our findings support the role of an evidence-based digital nutrition education program that is MS-specific to empower people with MS in their decision-making regarding dietary changes by acquiring new evidence-based knowledge.

Time availability and degree of social support appeared to influence the participants' ability to complete the program and make dietary changes. While fatigue is reported to be a common and debilitating symptom of MS [26], our findings did not indicate this had a detrimental impact on participants' ability to complete the program. Other research has reported that life circumstances that contribute to time-scarcity, such as family responsibilities and competing commitments, were barriers to making or maintaining lifestyle changes for people with MS [27]. Conversely, social support can assist in making changes to lifestyle habits [22, 27, 28]. Other online programs have reported that people with MS want to connect with peers through online discussion forums [25] to reduce feelings of isolation and enhance social connections [24, 29, 30], but have expressed similar frustration regarding low engagement within these forums [25]. Given the role of social support as an enabler for making and/or maintaining lifestyle changes, it is plausible that it may mitigate some of the detrimental effects of fatigue and time scarcity and assist people with MS in engaging with digital health interventions. Digital nutrition education programs could provide flexibility in the delivery mode to counteract time constraints, include family and/or friends as participants, and create engaging online discussion forums to foster a sense of community to support engagement, and completion of the program.

In our study, prioritising health and managing health concerns emerged as motivating factors to improve food choices and underscored the value placed on the nutrition education program for people who are newly diagnosed with MS. A diagnosis of MS is a critical event that spurs information-seeking regarding relevant, comprehensive, and trustworthy information on health-related behaviours specific to MS [24]. Furthermore, the exploration of participants' experiences with an MS online course revealed a consensus on the importance of learning about lifestyle modifications to better manage their MS and potentially improve their overall health [24]. Additionally, feedback from people with MS who participated in a pilot study of a digital health application highlighted the usefulness and relevance for those recently diagnosed [31]. These findings support our study, which indicates the potential benefits of a digital nutrition education program designed specifically to support dietary changes among people newly diagnosed with MS to improve their overall health.

The provision of physical resources, such as recipe booklets, practical activities, including goal setting and revising goals weekly, meal planning, modifying recipes, planning for fatigue days, combined with an online discussion forum for peer support, may assist people with MS in making dietary changes. Previous qualitative research has also highlighted a preference among people with MS for tangible resources including recipes and food lists, which also incorporate behaviour change techniques, such as goal setting and self-monitoring [32]. However, it is important to note that in line with our findings, these strategies were not universally applicable as some participants discussed their life circumstances that made certain activities less feasible. This underscores the importance of providing personally-relevant information and activities to support behaviour change in people with MS [29]. Therefore, nutrition education programs should include suitable behavioural supports such as tangible resources and practical activities tailored to people with MS to encourage program completion.

We used various techniques to ensure rigour and trustworthiness of this study, we included participants with a wide range of time since their MS diagnosis, and data collection continued until after saturation was reached [21, 33]. However, this study has some limitations. First, there is a potential for self-selection bias, whereby participants who responded to the email invitation may have been more motivated to complete the program and had a more positive response. We attempted to mitigate this by sending a second email detailing that we were seeking a range of perspectives, including those from people who did not complete the program, which was achieved. Second, audio interviews prohibited the interviewer from recognising and noting any nonverbal cues. Finally, the results are representative of the participants who took part in the feasibility study testing "Eating Well with MS", which only included participants from Australia, therefore, may not be representative of people with MS outside of Australia where there may be different access to dietetic services and dietary advice for MS.

### 4.2 Conclusion

This study offers insights into the capabilities, opportunities, and motivations of people with MS in Australia after engaging in a digital nutrition education program. Our findings reinforce the importance of MS-specific and evidence-based nutrition education to support people with MS to make dietary changes. Acquiring and/or validating nutrition knowledge, coupled with practical support mechanisms such as recipe booklets and goal-setting activities, emerged as crucial for facilitating engagement with the program and dietary behaviour change. Importantly, this study highlighted the role of life circumstances and social support in influencing program engagement. The next stage for program development is to explore further tailoring of the digital program to potentially mitigate some of the barriers to completion.

### 4.3 Practice Implications

Through targeting behaviour change in people with MS by addressing multiple aspects of capacity, opportunity, and motivation, nutrition education programs can be tailored to suit the MS community, and potentially improve their overall health and well-being through positive dietary changes. Healthcare professionals and program designers should consider flexible delivery modes of evidence-based content that addresses the specific needs and challenges faced by this population. Additionally, the importance of social support suggests that family, friend, or peer involvement could further facilitate participation and success.

We confirm all patient/personal identifiers have been removed or disguised so the patient/person(s) described are not identifiable and cannot be identified through the details of the story.